\documentclass[aps,twocolumn,nofootinbib,showpacs,prd]{revtex4}
\usepackage{graphicx}
\usepackage{amsfonts}
\usepackage{amssymb}
\usepackage{amsbsy}
\usepackage{amsmath}
\usepackage{latexsym}
\usepackage{natbib}
\usepackage{bm}
\usepackage{color}
\usepackage{float}

\begin{document}
\title{Self-Similar Collapse in Painlev\'e-Gullstrand Coordinates}
\author{Soumya Chakrabarti\footnote{soumya.chakrabarti@vit.ac.in}}
\affiliation{Department of Physics, School of Advanced Sciences, Vellore Institute of Technology \\ 
Vellore, Tiruvalam Rd, Katpadi, Tamil Nadu 632014 \\
India}
\author{Chiranjeeb Singha \footnote{chiranjeeb.singha@iucaa.in}}
\affiliation{Inter-University Centre for Astronomy and Astrophysics\\ Post Bag 4, Ganeshkhind, Pune - 411007\\ India}

\pacs{}

\date{\today}

\begin{abstract}
We report a family of self-similar exact solutions in General Relativity. The solutions are found in a Painleve-Gullstrand coordinate system but can also be transformed smoothly into a diagonal form. The solutions represent a gravitational collapse leading to three possible outcomes, depending on the parameter space : (i) a collapse followed by a bounce and dispersal of the clustered matter distribution, (ii) a rapid collapse followed by a bounce and an eventual re-collapse, and (iii) a standard collapse leading to zero proper volume. Profiles of the energy conditions are studied for all of the scenarios, and it is noted that a bounce is usually associated with a violation of the Null Energy Condition. It is found that more than one null surfaces (apparent horizons) can develop during the collapse. We also discuss that for a general metric tensor having a conformal symmetry, some regions of the parameter space allows a formation of null throat, much like a wormhole. Matching the metric with a Schwarzschild metric in Painleve-Gullstrand form leads to the geodesic equation for a zero energy falling particle in the exterior.
\end{abstract}

\maketitle

\section{Introduction}
A \textit{self-similar} object exhibits similar statistical properties on different length-scales. It is also understood as scale-invariance and one may refer to coast-lines drawn on a map or fractal patterns such as Koch snowflakes as practical examples. If a function (observable quantity) $u(x,t)$ has different values at different times but can be written as a function of $z = \frac{x}{t^{\alpha}}$, then the dimensionless quantity can exhibit a (dynamic) scale invariance and, in turn, an evolving self-similarity. As in elementary geometry, this is just a generalization of the idea of similar triangles. The sides of such triangles can change depending on the spatial coordinates, but the angle between the two arms (which is, a dimensionless quantity) remain the same. Self-similarity in general relativity (GR), or Riemannian geometry, is closely related to the idea of Killing symmetry, usually enforced on any metric tensor with a conformal Killing vector (CKV) $\eta$ obeying
\begin{equation}
L_{\eta}g_{ab}=\eta_{a;b}+\eta_{b;a} = \lambda g_{ab}.
\end{equation}

$\lambda$ can be a function of coordinates, however, we shall focus on a special class of CKV known as the homothetic killing vector (HKV) where $\lambda = 2$ (for a detailed discussion, see for instance \cite{maartens}). If a general relativistic metric tensor admits a homothetic killing vector $\xi$ satisfying $L_{\xi}g_{ab} = \xi_{a;b}+\xi_{b;a} = 2g_{ab}$, by a suitable transformation of coordinates all metric coefficients and dependent variables can be transformed into functions of a dimensionless combination of the space and the time coordinate \cite{scnb}. \\

In GR, remnants of this symmetry are likely to be found in large-scale structures. Although these structures seem static in nature, it is better to treat them as intermediate dynamical equilibrium phases of a general time-evolving process, such as a stellar collapse. \textit{Collapse} of massive stellar distributions has been a popular topic of discussion for almost a century. Most of our knowledge regarding this process are based on generalizations of the primary works by Oppenheimer and Snyder \cite{os}. While it can be proved that for an idealized (homogeneous, perfectly spherical) distribution of initially collapsing matter a gravitational collapse will always produce a zero proper volume, the same can't be said for a geometry exhibiting a special symmetry, such as self-similarity. Different phases of a collapse, such as the formation of a horizon or a singularity, also depend heavily on the background symmetry, as proved in a number of proposed models in literature \cite{literature}. Intuitively, this phenomenon should also have a connection with concepts of information paradox, as a \textit{complete} model of gravitational collapse should ideally portray a smooth evolution into small (Planck) length scales. A mechanism to allow quantum effects to generate enough modifications to drive something like a Hawking radiation \cite{ip} should also follow naturally. This would require the introduction of a quantum-corrected horizon and its time evolution, however, this has never been completely done till date. The closest candidate that can provide an analog model with a horizon to test quantum corrections on an appropriate scale is an \textit{Analog Black Hole}. Unruh pointed out that the equations governing sound waves propagating in an irrotational, barotropic fluid with negligible viscosity are the same as those for a massless scalar field, if an appropriate general relativistic metric can be found \cite{unruh1}. The curious implication is that sound horizons can behave similarly to event horizons in black holes and emit Hawking radiation. The early proposal was to try and detect Hawking radiation from sonic black holes, and this idea led to a simple static spherically symmetric solution of the Einstein field equations, written in Painlev\'e-Gullstrand-Lema\^itre (PGL) form \cite{unruh2, pg}. The PGL analog is often called a `River Model' due to the image it can portray: a flow of space with Newtonian escape velocity through a flat background. An event horizon develops whenever the flow velocity becomes equal to the speed of light (see \cite{river, riversc} for a detailed illustration). Therefore, an advantage of analog gravity framework is that working with a simple fluid flow one can portray (simulate) general relativistic dynamical evolutions, even on a length scale approaching the quantum limits. \\

We present a new exact solution of Einstein field equations in PGL coordinates. The solution can depict a spherically symmetric time-evolving gravitational collapse. The interior geometry preserves a self-similarity throughout the process. In literature, most of the self-similar solutions are found with different configurations of scalar field \cite{scnb, scalarcollapse}. These solutions often lead to the so-called \textit{critical phenomenon} characterized by a single parameter. Depending on the constraints in parameter space, the solution exhibits a transition between complete dispersal and black hole formation \cite{cp}. The solution we present is found for a collapsing sphere filled with a radiating imperfect fluid. Although, at the outset, there is no strict requirement for a scalar field, it must be mentioned that quite a few reasonable configurations of scalar field can mimic standard matter distributions such as dust or radiation \cite{matterscalar}. Therefore, the solution found here seems applicable in general. The transformation from a diagonal metric tensor into a PGL form enforces a few additional constraints. For instance, any stationary black hole metric must be spatially flat at any fixed time, up to a conformal factor \cite{riversc}, to admit a consistent PGL analog. For a self-similar collapse, we incorporate a time evolution in this conformal factor and also make it a function of $z = \frac{t}{r}$. For a smooth transition into PGL form, the metric components must obey a set of three differential equations. We derive and solve them to find an exact solution. \\

Section $II$ includes the mathematical setup, detailed methodology, the exact solution and a discussion on the nature of the solutions. In section $III$ a discussion on the validity of the energy conditions is included. The conditions for forming a null surface during this self-similar collapse are discussed in Section $IV$. In Section $V$, the smooth matching with a vacuum exterior is discussed briefly, and the article is concluded in Section $VI$.

\section{An exact self-similar solution}

\begin{figure}[t]
\begin{center}
	\includegraphics[width=\linewidth]{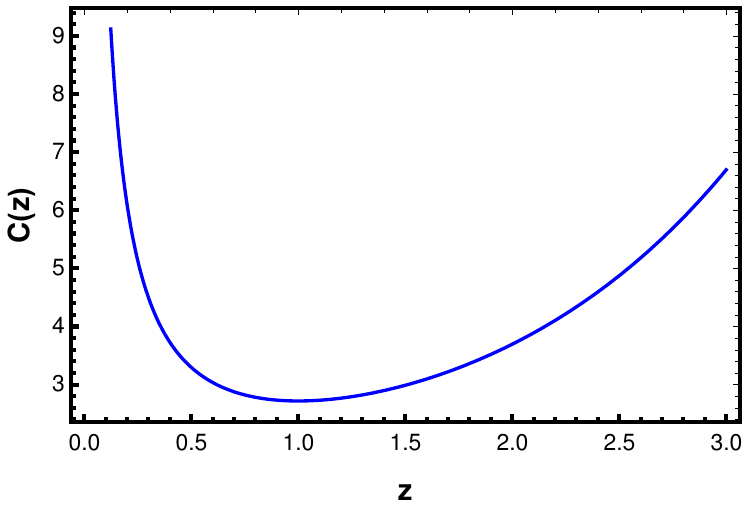}
	\includegraphics[width=\linewidth]{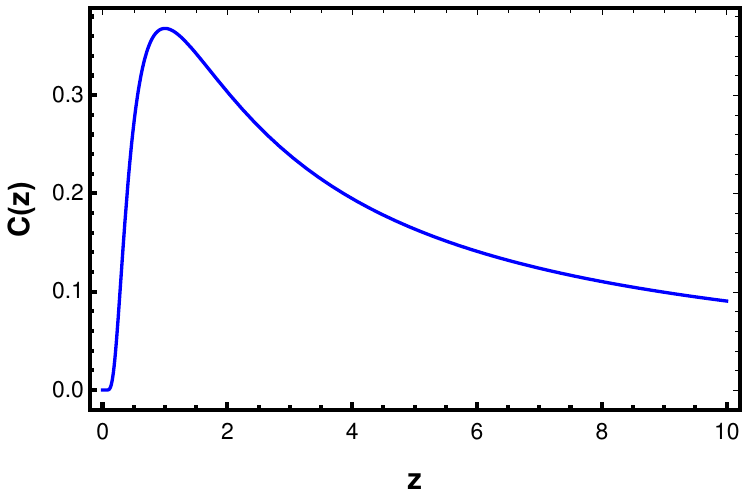}
	\includegraphics[width=\linewidth]{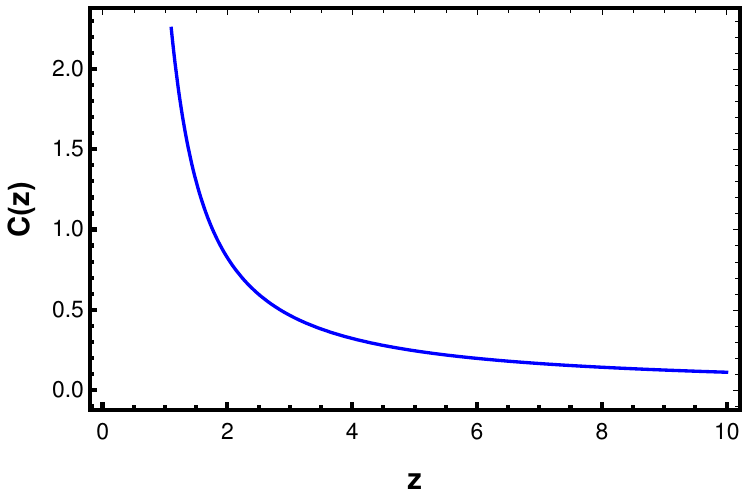}	
\caption{Radius of two-sphere as a function of z : Top $\Rightarrow$ $\frac{e^{-\frac{z^{-n}}{n}}}{z}$, $n < 0$. Middle  $\Rightarrow$ $\frac{e^{-\frac{z^{-n}}{n}}}{z}$, $n > 0$. Bottom : $\Rightarrow$ $\frac{e^{\frac{z^{-n}}{n}}}{z}$.}
\label{solution}
\end{center}
\end{figure}

We start with a general spherically symmetric spacetime metric, written in the form
\begin{equation}\label{startmet}
ds^{2}_{-} = C^2\left(A^2 dt^2 - B^2 dr_{c}^2 - r_{c}^2 d\Omega^2\right).
\end{equation}
It describes an interior geometry filled with a locally anisotropic fluid expressed through the energy-momentum tensor
\begin{equation}
T_{\alpha\beta}=(\rho+p_{t})u_{\alpha}u_{\beta}-p_{t}g_{\alpha\beta}+ (p_r-p_{t})\chi_{\alpha} \chi_{\beta}+q_{\alpha}u_{\beta}+q_{\beta}u_{\alpha}.
\end{equation}
$\rho$, $p_{t}$ and $p_r$ are density, tangential and radial pressure. $q^{\alpha} = (0,q,0,0)$ is the radial heat flux. The four-velocity and the unit four-vector in radial direction follow usual normalizations
\begin{equation}
\label{norm}
u^{\alpha}u_{\alpha}=1,\quad\chi^{\alpha}\chi_{\alpha}=-1,\quad\chi^{\alpha}u_{\alpha} = 0.
\end{equation}
We introduce the transformation $r_{c}C = r$ to write
\begin{equation}\label{transformation}
dr_{c} = \frac{1}{C}(dr-r_{c}dC) = \frac{1}{C}\left\lbrace dr - r_{c}(\dot{C}dt + C'dr)\right\rbrace.
\end{equation}
A dot represents a derivative with respect to $t$, and a prime is a derivative with respect to $r$. The transformation allows us to write Eq. (\ref{startmet}) as
\begin{equation}
ds^{2}_{-} = A^2 C^2 dt^2 - B^2\left\lbrace dr - \frac{r}{C}(\dot{C}dt + C'dr)\right\rbrace^{2} - r^{2}d\Omega^{2}.
\end{equation}
After simplification, it can be written as
\begin{eqnarray}\label{interior}\nonumber
&&ds^{2}_{-} =C^2\Bigg(A^{2} - \frac{r^{2}B^{2}\dot{C}^{2}}{C^{4}}\Bigg)dt^{2} + \Bigg(\frac{2rB^{2}\dot{C}}{C} \\&&\nonumber
- \frac{2r^{2}B^{2}\dot{C}C'}{C^{2}}\Bigg) dtdr - \Bigg(B^2 - \frac{2rB^{2}C'}{C} + \frac{r^{2}B^{2}C'^{2}}{C^{2}}\Bigg)dr^{2} \\&&
- r^{2}d\Omega^{2}.
\end{eqnarray}

Comparing term-by-term with a generic PGL metric
\begin{eqnarray}\label{gen_PG}
ds^2 = (1-\zeta^2) dt^2 \pm 2\zeta drdt - dr^2 - r^2 d\Omega^{2},
\end{eqnarray}
we deduce that in order to have a smooth PGL analogue, the original metric components and $\zeta(r,t)$ should satisfy the following set of differential equations
\begin{eqnarray}\label{formationeq}
&& \pm 2\zeta C^2 = \frac{2rB^{2}\dot{C}}{C} - \frac{2r^{2}B^{2}\dot{C}C'}{C^{2}},\nonumber \\&&
(1-\zeta^2) C^2 = A^{2} C^2 - \frac{r^{2}B^{2}\dot{C}^{2}}{C^{2}}, \nonumber\\&&
B^2 - \frac{2rB^{2}C'}{C} + \frac{r^{2}B^{2}C'^{2}}{C^{2}} = C^2.
\end{eqnarray}

Here, $C$ is a function of $t$ and $r$. If we choose the metric coefficients to be self-similar, i.e., functions of $z = \frac{t}{r}$, Eq. (\ref{formationeq}) becomes,
\begin{eqnarray}
&& \pm 2\zeta C^2 = \frac{2B^{2}}{C}\bigg(\frac{d C}{d z}\bigg) + \frac{2 B^{2} z}{C^{2}}\bigg(\frac{d C}{d z}\bigg)^2,\label{formationeqz1} \\&&
(1-\zeta^2) C^2 = A^{2} C^2 - \frac{B^{2}}{C^{2}}\bigg(\frac{d C}{d z}\bigg)^2, \label{formationeqz2}\\&&
B^2 + \frac{2zB^{2}}{C}\bigg(\frac{d C}{d z}\bigg) + \frac{B^{2}z^{2}}{C^{2}}\bigg(\frac{d C}{d z}\bigg)^2 = C^2.\label{formationeqz3}
\end{eqnarray}

Using the ansatz $B = z^{n} C$ we find from Eq. (\ref{formationeqz1})
\begin{eqnarray}\label{formationeqz}
C = \frac{C_{1}}{z}~e^{-z^{-n}/n},\; \textit{or}\; C= \frac{C_{1}}{z}~e^{z^{-n}/n},
\end{eqnarray}

where $C_{1}$ is a constant of integration. In Fig. \ref{solution}, we see three scenarios : (i) a collapse and bounce which avoids any formation of singularity ($C=\frac{e^{-\frac{z^{-n}}{n}}}{z}$, $n < 0$), (ii) a rapid collapse and bounce followed by a recollapse ($C = \frac{e^{-\frac{z^{-n}}{n}}}{z}$, $n > 0$) and (iii) a standard collapsing scenario ($\frac{e^{\frac{z^{-n}}{n}}}{z}$). For $C = \big(C_{1}e^{-z^{-n}/n}\big)/z$, the other metric coefficients are found from Eqs. (\ref{formationeqz2} and \ref{formationeqz3}) as
\begin{equation}
\pm\zeta = \big(1- z^{n}\big)/z, ~~ A^{2} = 1,
\end{equation}
Similarly, For $C = \big(C_{1} e^{z^{-n}/n}\big)/z$, the coefficients can be found from Eqs. (\ref{formationeqz2} and \ref{formationeqz3}) as

\begin{equation}
\pm\zeta = \big(1+ z^{n}\big)/z, ~~ A^{2} = 1.
\end{equation}

\begin{figure}[H]
\begin{center}
	\includegraphics[width=\linewidth]{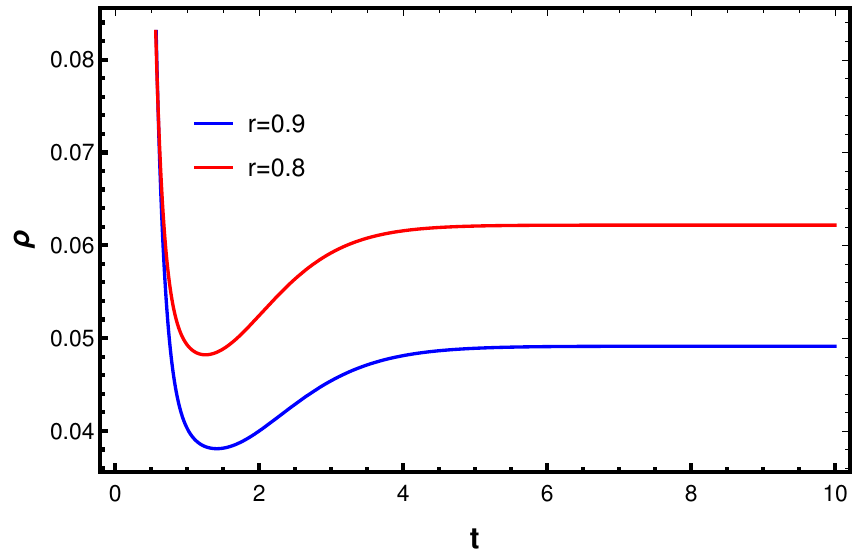}
	\includegraphics[width=\linewidth]{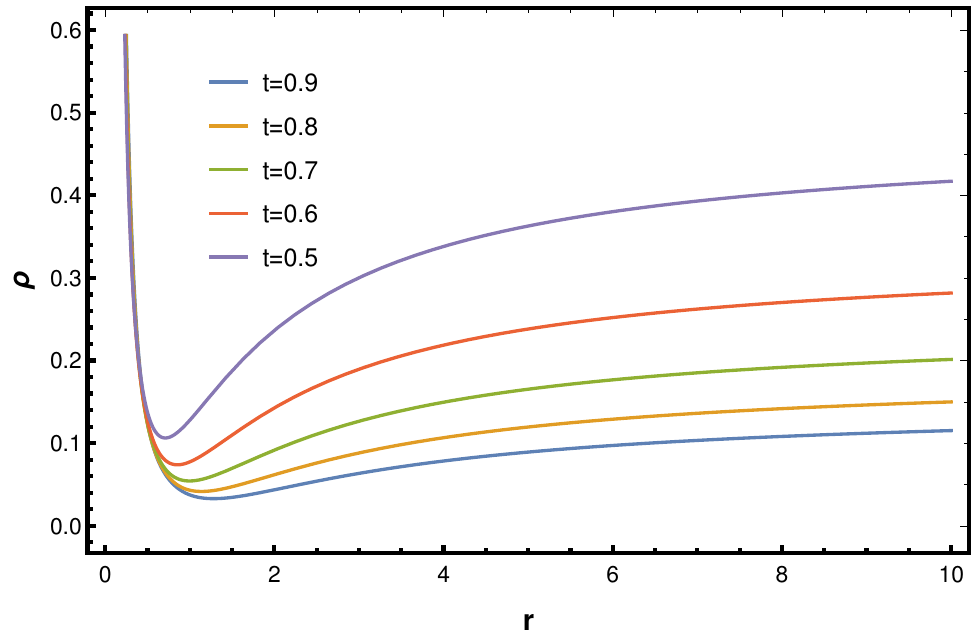}
	\includegraphics[width=\linewidth]{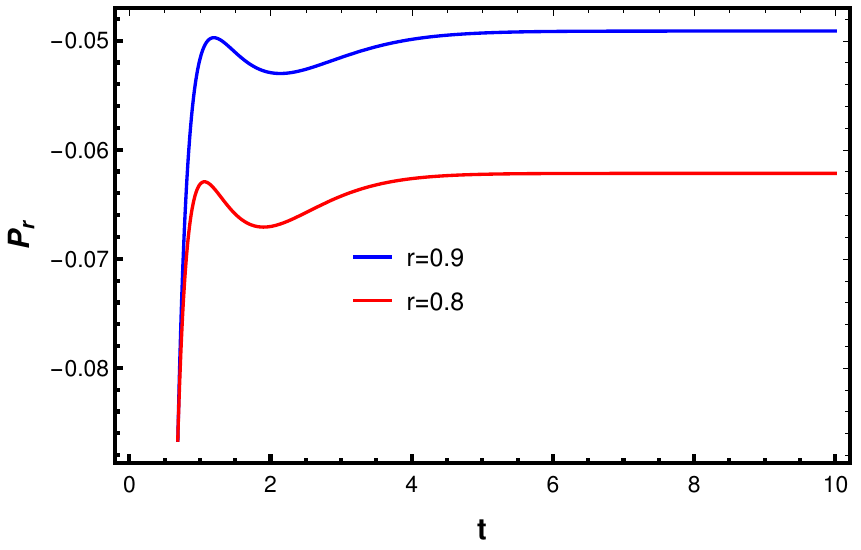}	
	\includegraphics[width=\linewidth]{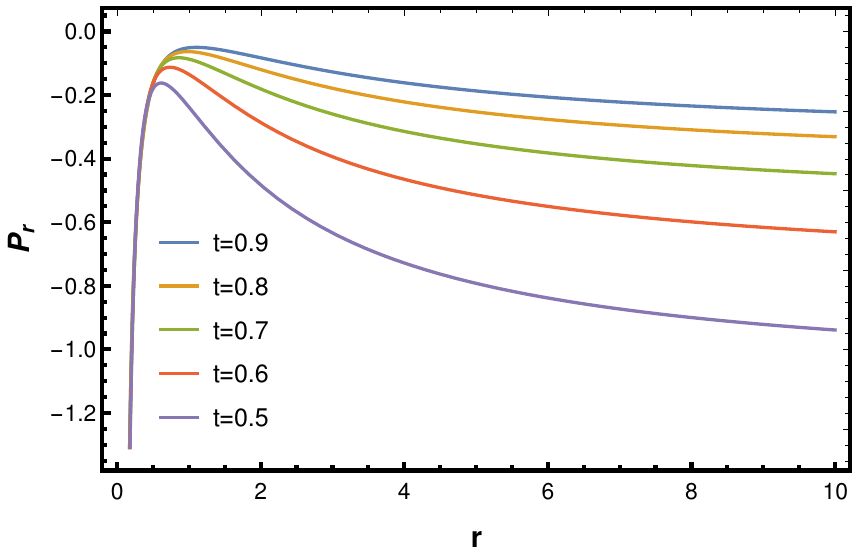}
\caption{Evolution of density $(\rho)$ and radial pressure $(P_{r})$ with respect to time $t$ for different collapsing shells (labeled by $r$) and as a function of $r$ for different values of $t$. Radius of two-sphere is taken as $\frac{e^{-\frac{z^{-n}}{n}}}{z}$, $C_{1}= G =1$ and $n = -1$.}
\label{density_radp}
\end{center}
\end{figure}

\section{Energy Momentum Distribution}

These solutions are found simply from the requirement for a smooth transformation of a diagonal and a non-diagonal metric tensor. We can treat the rest of the field equations as constraints by pressing this advantage of working in the so-called river frame. It can also be checked that the Ricci and Kretschmann scalars diverge at $r \rightarrow 0$. To use the field equation $G_{\mu\nu} = 8 \pi G T_{\mu\nu}$ we write the nonzero components of Einstein tensor as

\begin{eqnarray}
&& G^0_{\;0} = -\frac{2\zeta\zeta^{\prime}}{r} - \frac{\zeta^2}{r^2},\\&&
G^1_{\;1} = -\frac{2\zeta\zeta^{\prime}}{r} - \frac{\zeta^2}{r^2} - \frac{2\dot{\zeta}}{r}, \\&&
G^1_{\;0} = \frac{2\zeta\dot{\zeta}}{r}, \\&&
G^2_{\;2} = G^3_{\;3} = -\frac{\dot{\zeta}+2\zeta\zeta^{\prime}}{r} - \dot{\zeta}^{\prime} - \zeta\zeta^{\prime\prime} - \zeta^{\prime\,2}. \label{FE_gen}
\end{eqnarray}

\begin{figure}[H]
\begin{center}
	\includegraphics[width=\linewidth]{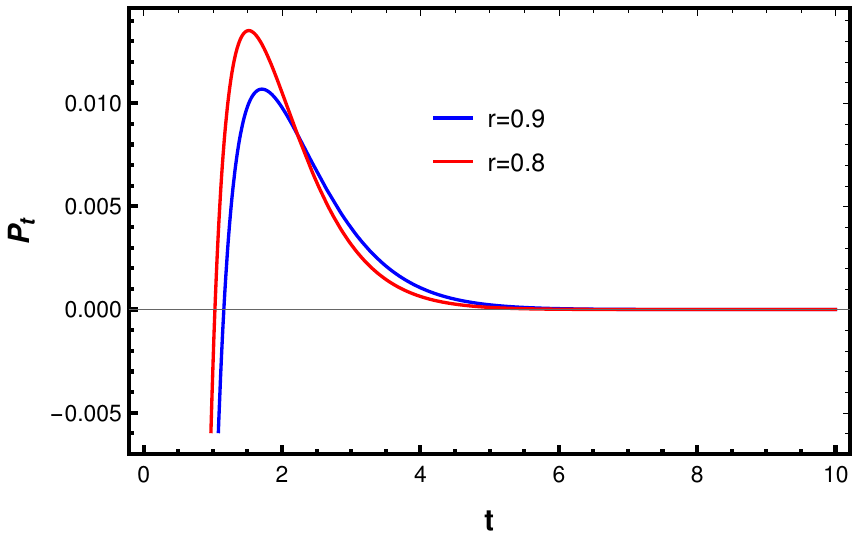}
	\includegraphics[width=\linewidth]{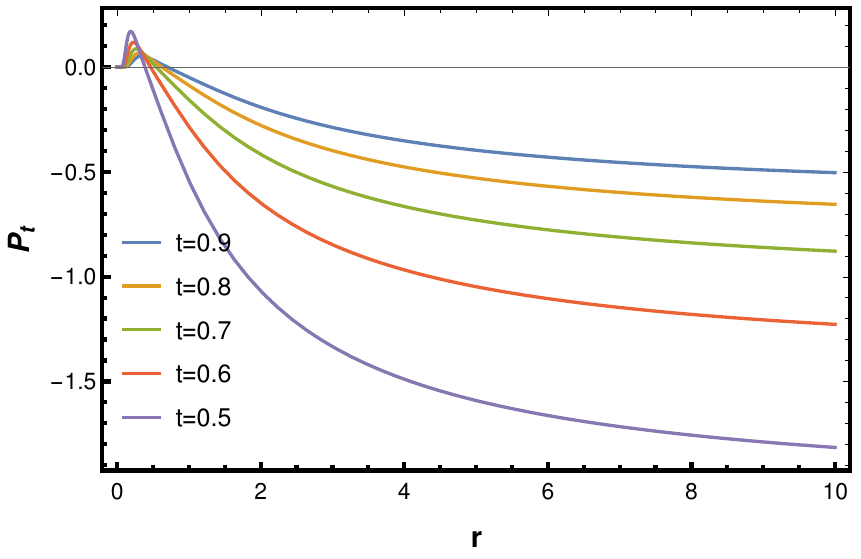}
	\includegraphics[width=\linewidth]{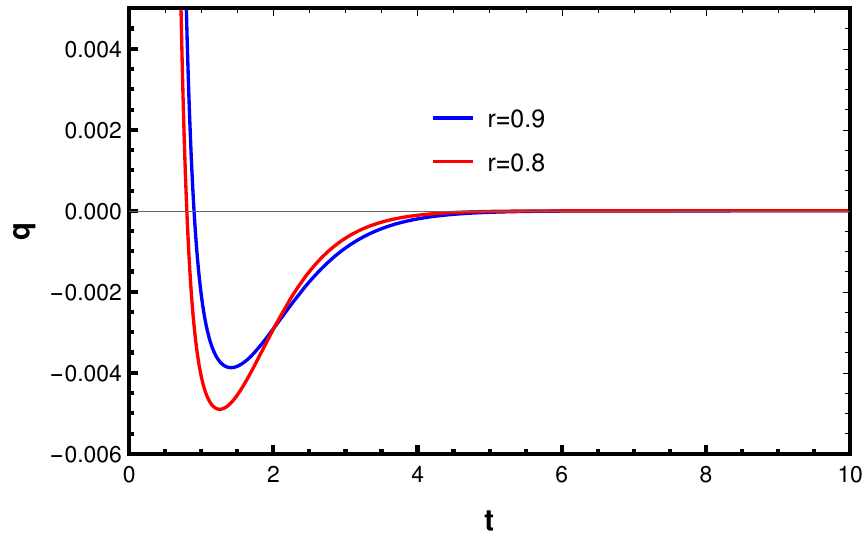}
	\includegraphics[width=\linewidth]{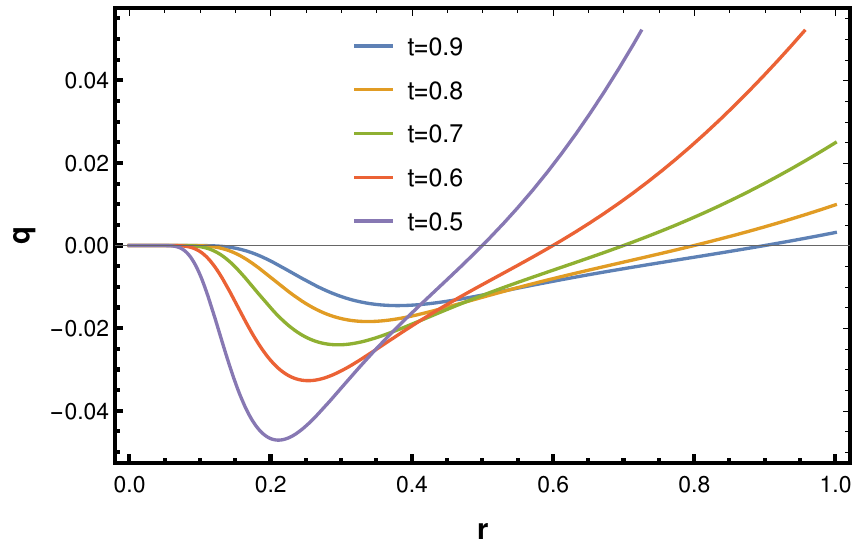}
\caption{Evolution of tangential pressure and heat flux with respect to time $t$ for different collapsing shells (labeled by $r$) and as a function of $r$ for different values of $t$. Radius of two-sphere is taken as $\frac{e^{-\frac{z^{-n}}{n}}}{z}$, $C_{1}= G =1$ and $n = -1$.}
\label{tanp_flux}
\end{center}
\end{figure}

We do not write the explicit expressions of these components for the sake of brevity. However, the qualitative nature of what they represent is shown through graphs. In Figs. \ref{density_radp} and \ref{tanp_flux}, we plot the energy-momentum tensor components for a collapse and bounce, i.e., $C = \frac{e^{-\frac{z^{-n}}{n}}}{z}$ with $n < 0$. For all values of time and for all collapsing shells (labeled by specific values of $r$), $\rho\ge 0$ and, therefore, weak energy condition is satisfied. We plot the density profile as a function of time for different collapsing shells (top), as well as a radial profile (second from the top) for different snapshots of time values. The same is done for radial pressure, whose time evolution is shown in the graph third from the top, while the radial profile is shown in the graph below. Radial pressure is always negative. Its modulus increases with the collapse before starting to decay, along with the bounce. The tangential pressure is negative when the collapse begins. It switches over to positive values once the nature of the initially collapsing sphere changes into a bouncing phase. Eventually, the tangential pressure decays to zero over time. The heat flux profile is exactly the opposite: it is positive during the collapse and has a transition into negative values with the bounce. The flux also dies down to zero with time, signaling a dispersion of matter energy-momentum distribution. Surprisingly, the density decays for a time domain due to a large non-zero outgoing heat flux. Knowing the components of a general relativistic energy-momentum tensor gives us an advantage: to investigate the validity of energy conditions. The conditions are usually derived from the eigenvalue equation of the tensor, written in matrix form \cite{ec}. Amongst these, the Null Energy Condition (NEC) is the most important since, in GR, it is directly related to the Null Convergence Condition.

It dictates whether a family of null geodesics shall collapse to singularity or diverge away from one another. Mathematically, the condition holds if for all null vectors $T_{\mu\nu} k^{\mu} k^{\nu} \geq 0$. The weak energy condition ensures that the energy density component is always non-negative. The strong energy condition (SEC) is satisfied if for any timelike unit vector $w^{\alpha}$, $2 T_{\alpha\beta} w^{\alpha} w^{\beta} + T \geq 0$, where $T$ is the trace of the energy-momentum tensor. SEC is only violated if the energy density is negative or if a large negative pressure component of the energy-momentum tensor exists. For a detailed analysis, we refer to the work by Kolassis, Santos, and Tsoubelis, Pimentel, Lora-Clavijo, and Gonzalez. \\

\begin{figure}[H]
\begin{center}
	\includegraphics[width=\linewidth]{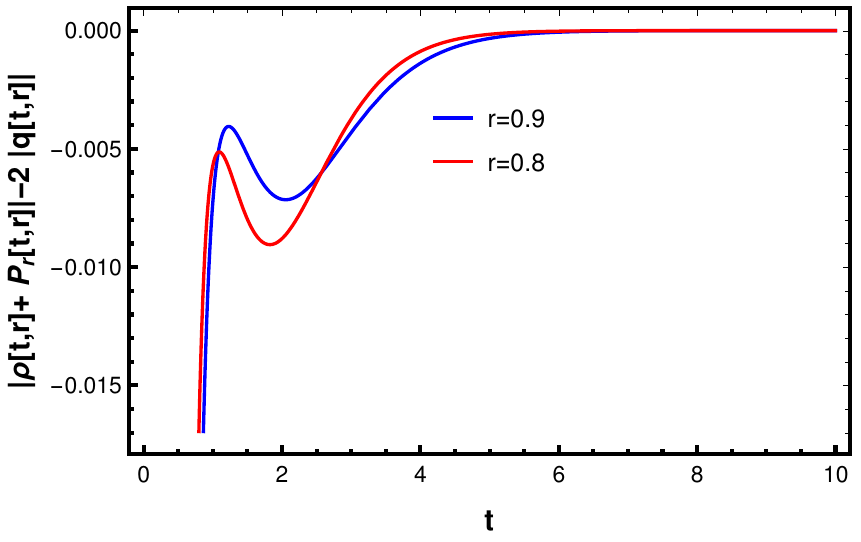}
	\includegraphics[width=\linewidth]{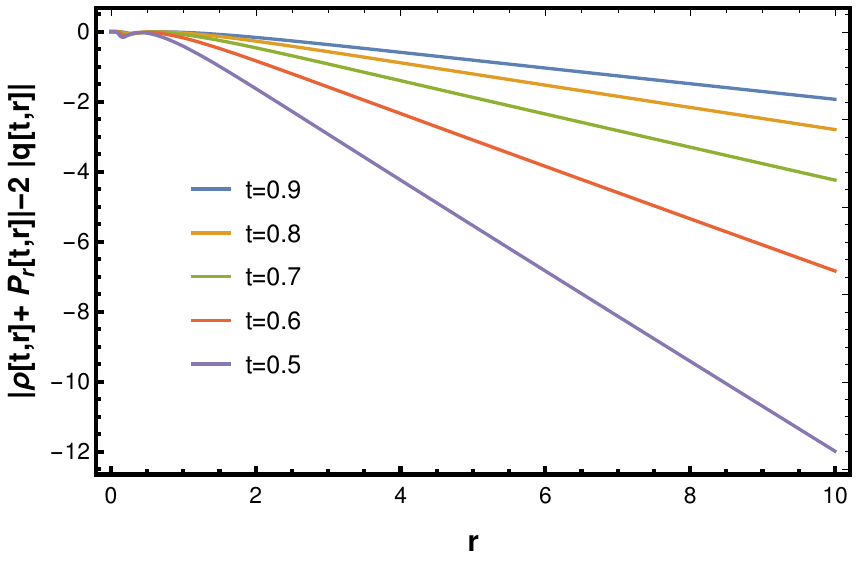}
	\includegraphics[width=\linewidth]{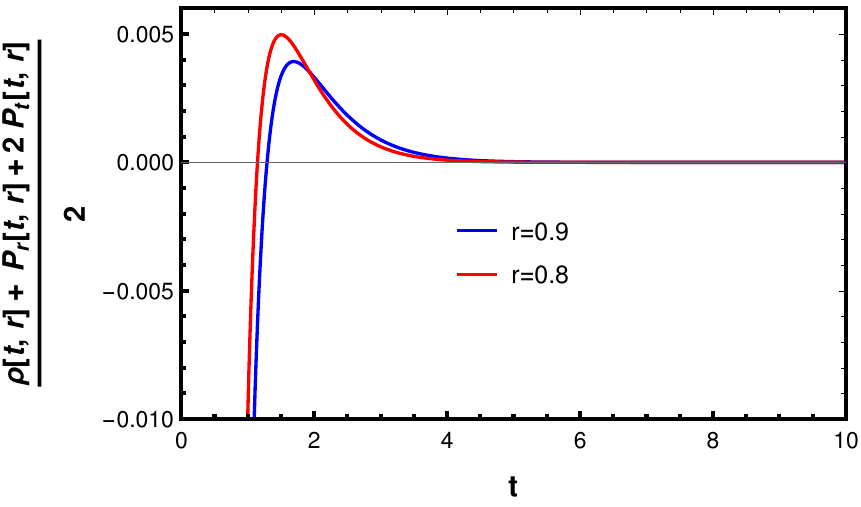}
	\includegraphics[width=\linewidth]{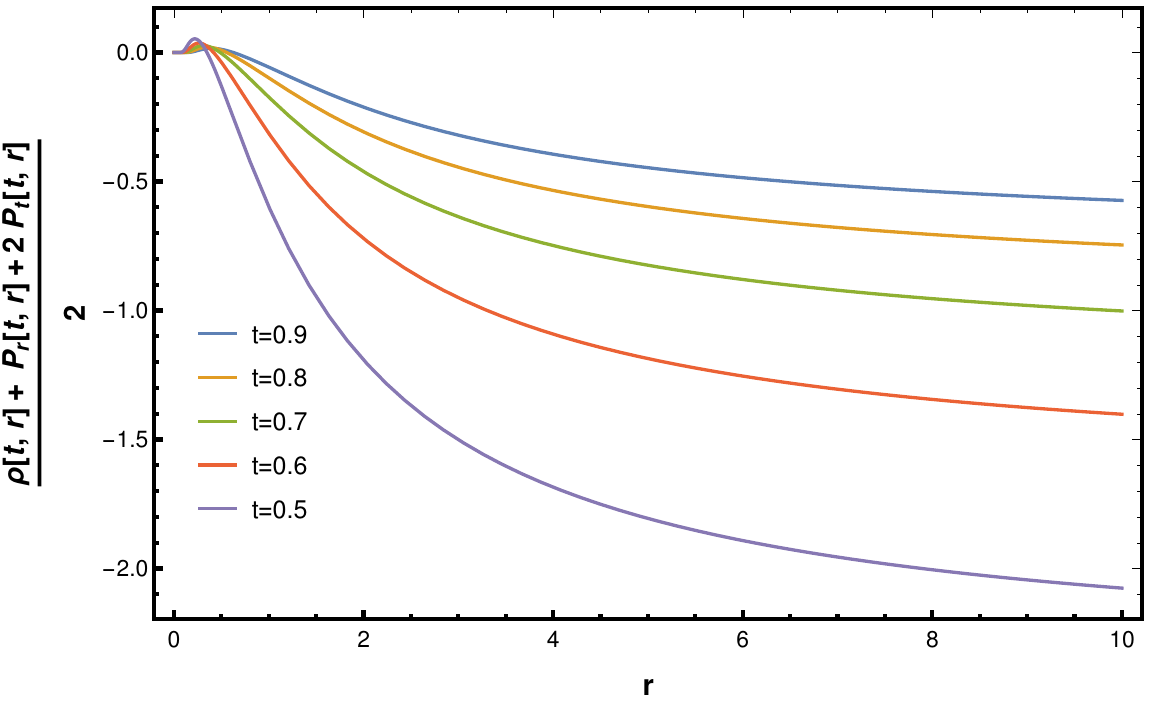}
	\caption{Evolution of Null Energy Condition and Strong Energy Condition with respect to time $t$ for different collapsing shells (labeled by $r$) and as a function of $r$ for different values of $t$. Radius of two-sphere is taken as $\frac{e^{-\frac{z^{-n}}{n}}}{z}$, $C_{1}= G =1$ and $n = -1$.}
\label{NEC_SEC}
\end{center}
\end{figure}

The profiles for NEC and SEC are shown in Fig. \ref{NEC_SEC}. NEC, as a function of time, is given in the graph on top for two different $r$-values, i.e., two collapsing shells. We also give the radial profile (second from the top) of NEC for different snapshots of time values. The same is done for SEC, whose time evolution is shown in the graph third from the top, and the radial profile is shown in the graph below. While the weak energy condition is never violated ($\rho > 0$), it is evident that both the NEC and SEC are violated during the bounce. It is quite possible that this violation generates an effective negative pressure and dominant heat flux, resulting in an eventual dispersion (and an aversion to singularity formation). \\

\begin{figure}[H]
\begin{center}
	\includegraphics[width=\linewidth]{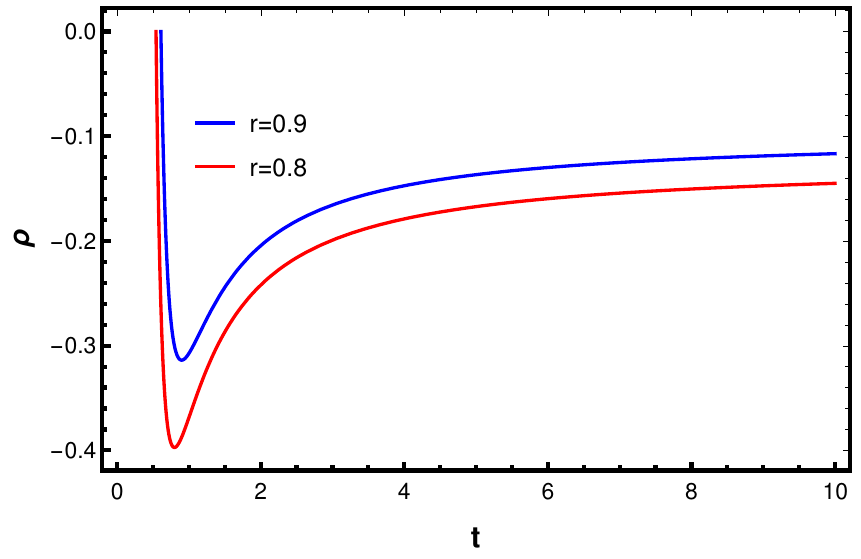}
	\includegraphics[width=\linewidth]{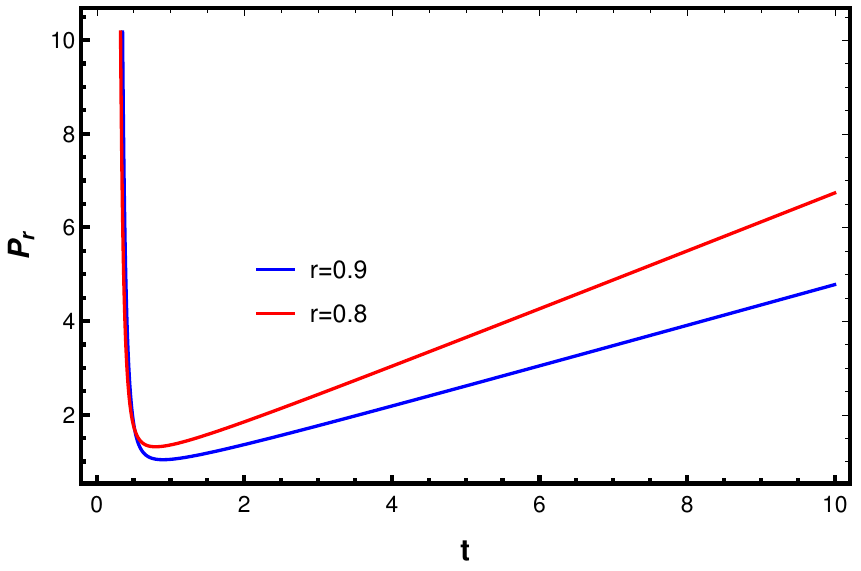}
	\includegraphics[width=\linewidth]{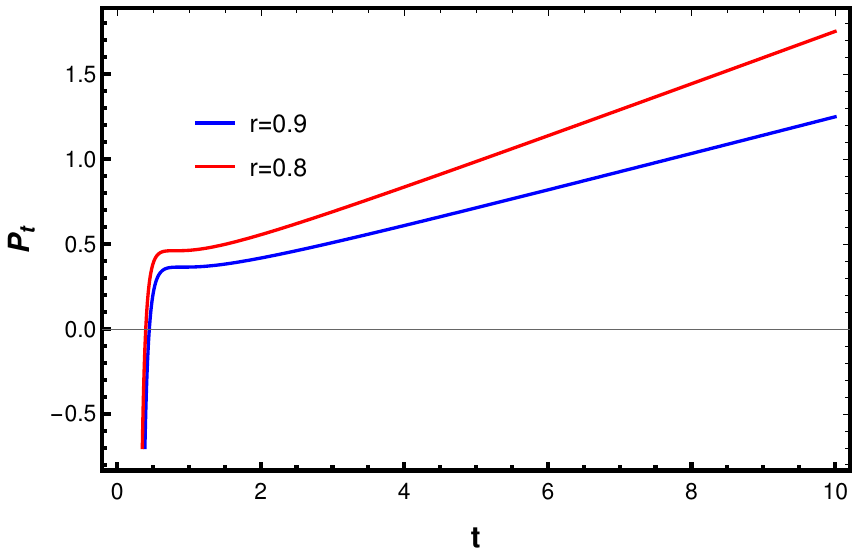}
	\includegraphics[width=\linewidth]{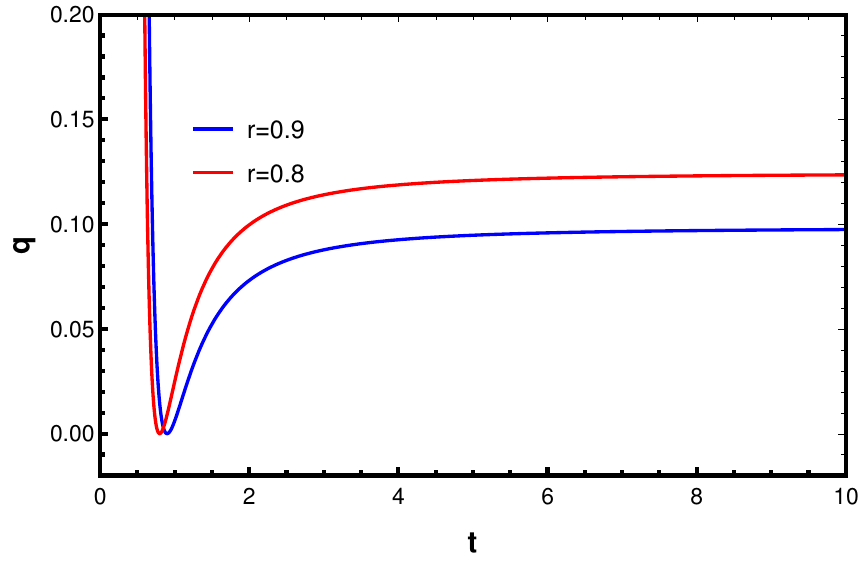}
\caption{Evolution of density, radial pressure, tangential pressure, and heat flux with respect to time $t$ for different collapsing shells (labeled by $r$) for the case of collapse-bounce-recollapse. Radius of two-sphere is taken as $\frac{e^{-\frac{z^{-n}}{n}}}{z}$, $C_{1}= G =1$ and $n = 1$.}
	\label{em_components_bounce_2}
\end{center}
\end{figure}

\begin{figure}[H]
\begin{center}
	\includegraphics[width=\linewidth]{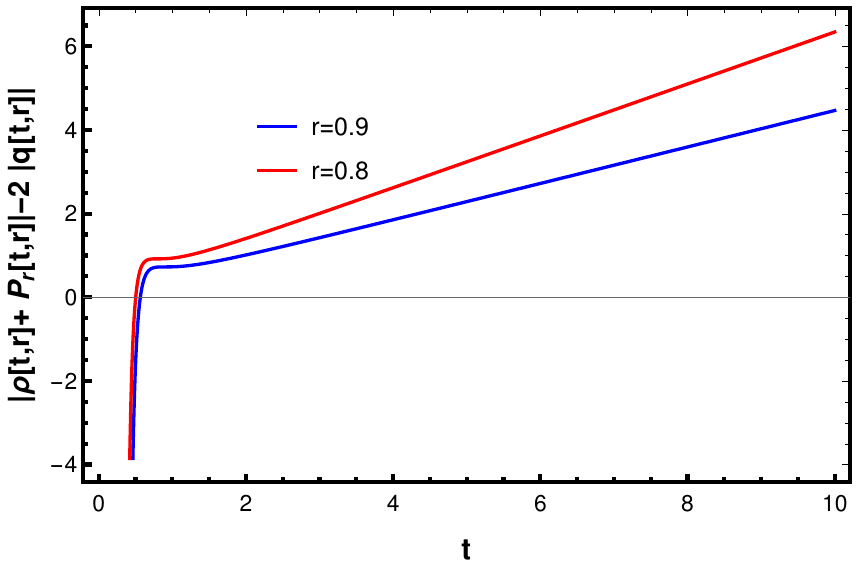}
	\includegraphics[width=\linewidth]{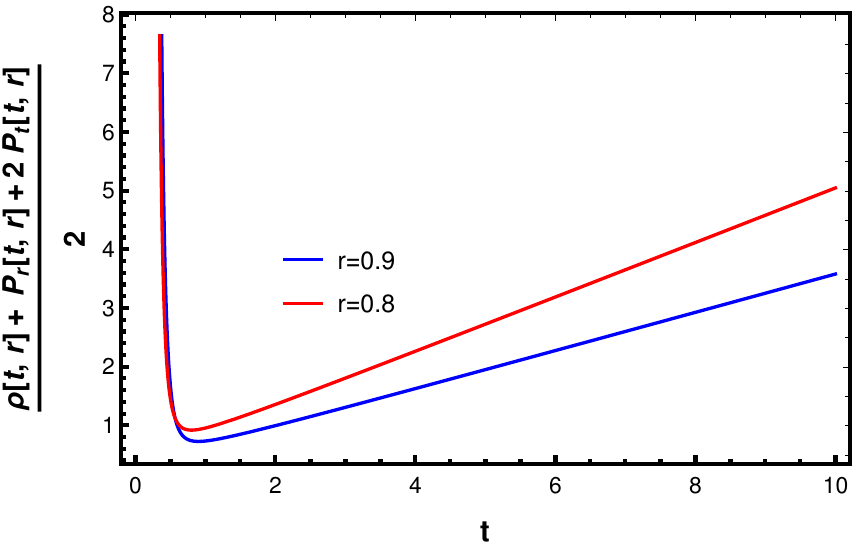}
\caption{Evolution of Null Energy Condition and Strong Energy Condition with respect to time $t$ for different collapsing shells (labeled by $r$) for the case of collapse-bounce-recollapse. Radius of two-sphere is taken as $\frac{e^{-\frac{z^{-n}}{n}}}{z}$, $C_{1}= G =1$ and $n = 1$.}
	\label{NEC_SEC_positive}
\end{center}
\end{figure}

We recall that for $C = \frac{e^{-\frac{z^{-n}}{n}}}{z}$, $n > 0$, we saw the sphere evolving through a collapse-bounce-recollapse phase. We show the energy-momentum tensor components as a time function in Fig. \ref{em_components_bounce_2}. Different curves in each of the graphs signify different collapsing shells. The notable difference compared to the first example of simple bounce is the fact that for all collapsing shells, $\rho \leq 0$, which signals the breakdown of weak energy condition and disfavors the choice of $C = \frac{e^{-\frac{z^{-n}}{n}}}{z}$ with $n > 0$. The other components, i.e., radial pressure, tangential pressure, and heat flux profile, are positive. Moreover, as Fig. \ref{NEC_SEC_positive} suggests, both the NEC and SEC are satisfied. Nevertheless, the origin of an outright negative energy density is difficult to explain, and this example can be treated as a toy model.

\begin{figure}[H]
\begin{center}
	\includegraphics[width=\linewidth]{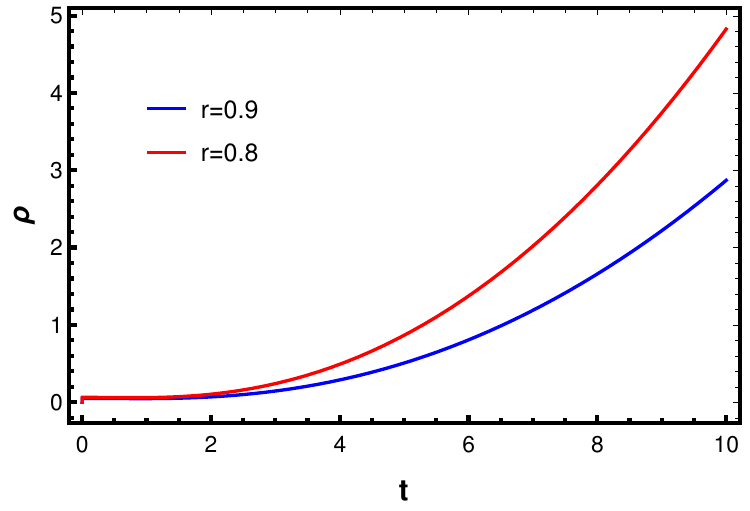}
	\includegraphics[width=\linewidth]{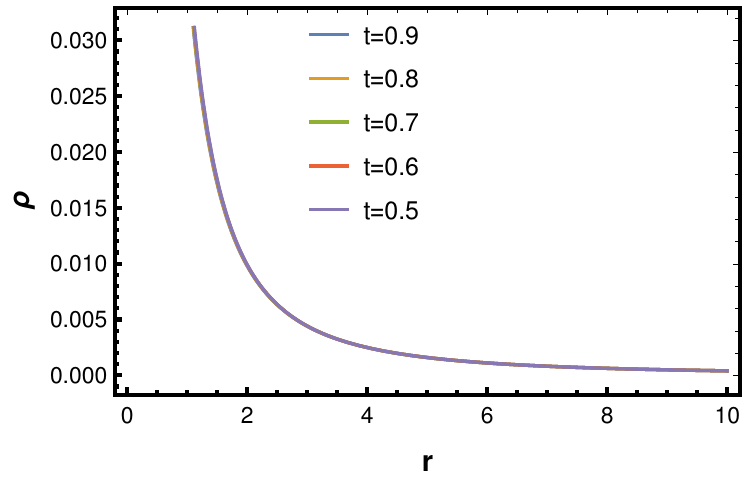}
	\includegraphics[width=\linewidth]{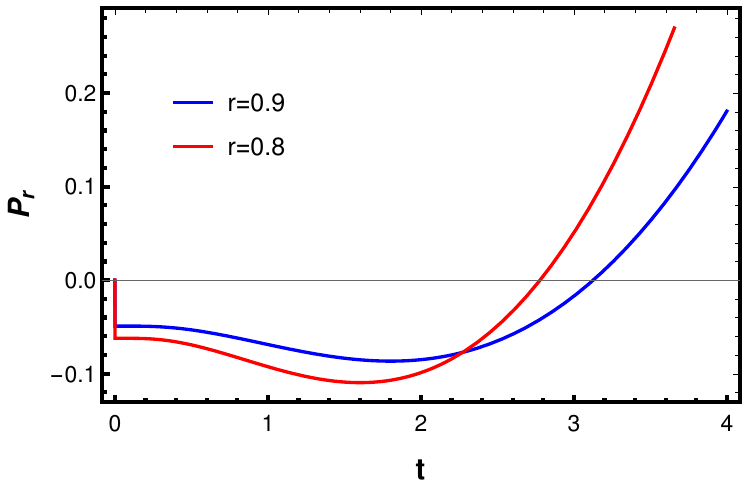}
	\includegraphics[width=\linewidth]{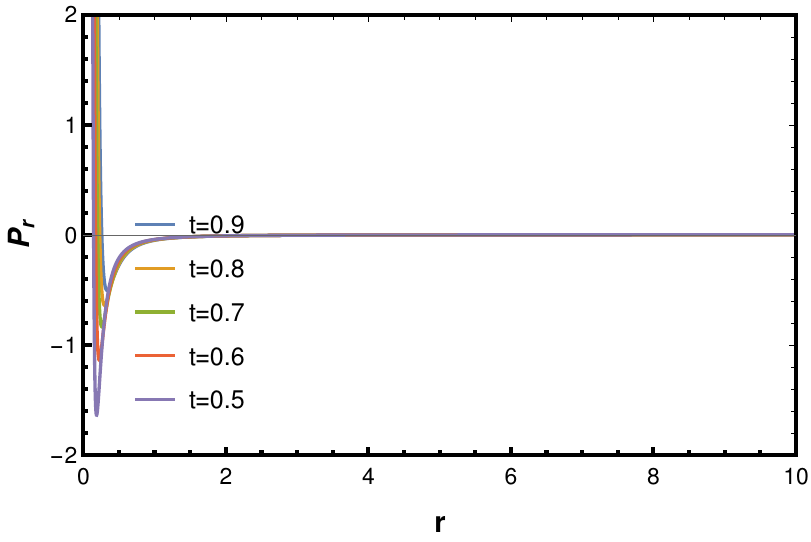}
\caption{Evolution of density $(\rho)$ and radial pressure $(P_{r})$ with respect to time $t$ for different collapsing shells (labeled by $r$) and as a function of $r$ for different values of $t$. Radius of two-sphere is taken as $\frac{e^{\frac{z^{-n}}{n}}}{z}$, $C_{1}= G =1$ and $n = 1$.}
\label{density_radp_1}
\end{center}
\end{figure}

\begin{figure}[H]
\begin{center}
	\includegraphics[width=\linewidth]{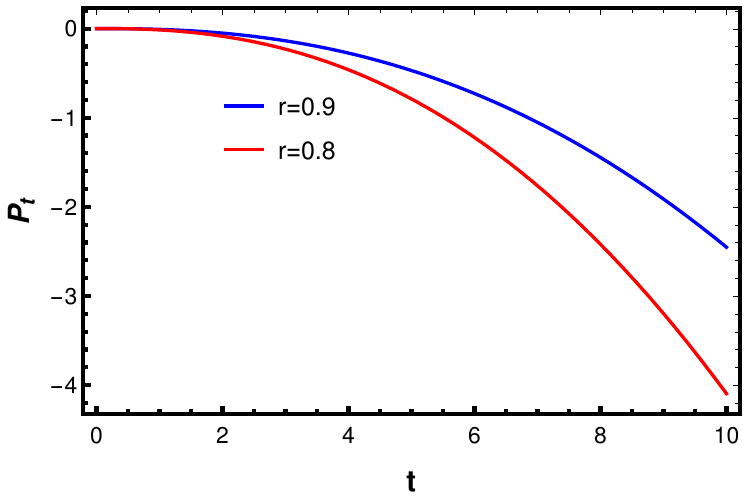}
	\includegraphics[width=\linewidth]{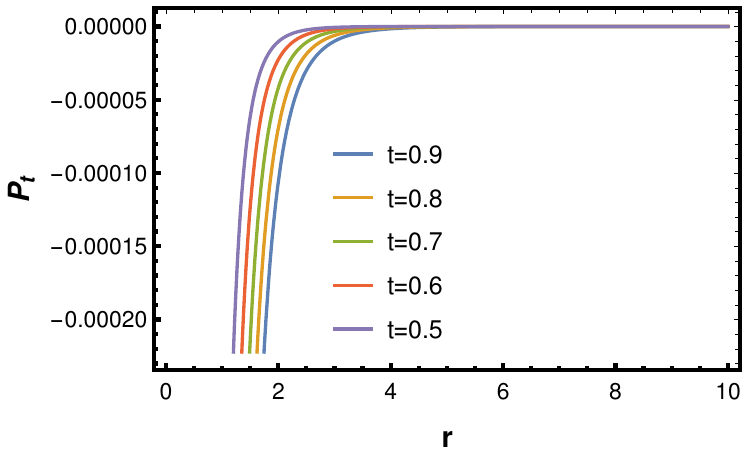}
	\includegraphics[width=\linewidth]{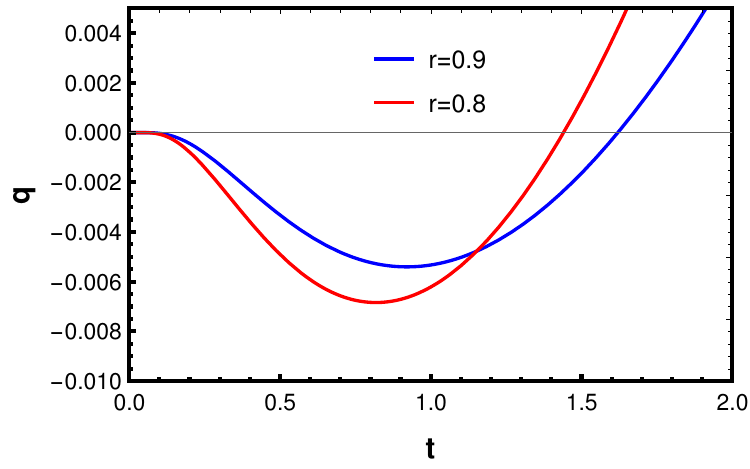}
	\includegraphics[width=\linewidth]{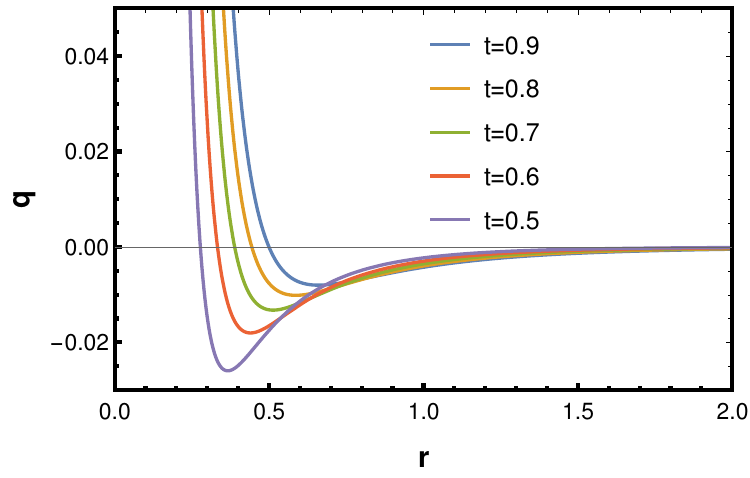}
\caption{Evolution of tangential pressure and heat flux with respect to time $t$ for different collapsing shells (labeled by $r$) and as a function of $r$ for different values of $t$. Radius of two-sphere is taken as $\frac{e^{\frac{z^{-n}}{n}}}{z}$, $C_{1}= G =1$ and $n = 1$.}
\label{tanp_flux_1}
\end{center}
\end{figure}

\begin{figure}[H]
\begin{center}
	\includegraphics[width=\linewidth]{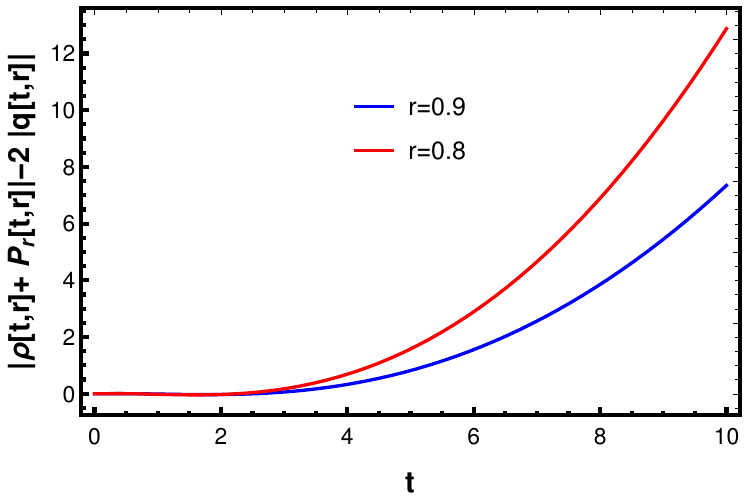}
	\includegraphics[width=\linewidth]{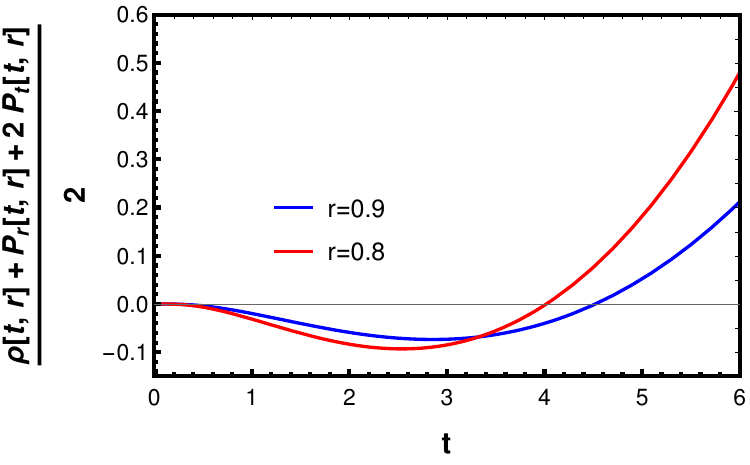}
\caption{Evolution of Null Energy Condition and Strong Energy Condition with respect to time $t$ for different collapsing shells (labeled by $r$). Radius of two-sphere is taken as $\frac{e^{\frac{z^{-n}}{n}}}{z}$, $C_{1}= G =1$ and $n = 1$.}
	\label{NEC_SEC_1}
\end{center}
\end{figure}

Finally, we follow the defined road for the standard collapsing scenario where the radius of the two-sphere is written as $\frac{e^{\frac{z^{-n}}{n}}}{z}$. We emphasize the fact that for this choice, both negative and positive values of $n$ produce a standard collapsing geometry. Once again, we do not write the explicit expressions of energy-momentum components but represent the evolutions in graphs. In Figs. \ref{density_radp_1} and \ref{tanp_flux_1}, we plot the energy-momentum tensor components for a self-similar collapsing fluid. For all values of time and for all collapsing shells (labeled by specific values of $r$), $\rho \geq 0$ and therefore, weak energy condition is always satisfied. We plot the density profile as a function of time for different collapsing shells (top), as well as a radial profile (second from the top) for different snapshots of time values. There is a gradual build-up of density when near the central shell, i.e., around $r \sim 0$. The same is done for radial pressure, whose time evolution is shown in the graph third from the top, and the radial profile is shown in the graph below. Radial pressure is initially negative. However, there is a crossover into the positive domain during the course of collapse. The large, positive build-up in radial pressure is realized near the central collapsing shells. The tangential pressure is negligible when the collapse begins. However, it is a matter of intrigue that this component starts to grow along with the collapse. This can be a by-product of the symmetry in space-time, i.e., the self-similarity. Quite similarly, the initial heat flux is negligible. It grows into negative values with the collapse, followed by a crossover into the positive domain once the shells start falling into a singularity. The inner-most shells contribute most to the total radiation produced during the collapse, and as a result, near the central shell, the heat flux tends to go to infinity. As one moves towards outer shells, there is negligible heat flux. The profiles for NEC and SEC are shown in Fig. \ref{NEC_SEC_1}. NEC is given in the graph on top for two different $r$-values, i.e., two collapsing shells, and suggests that there is no violation. The SEC is violated for a while during the initial phases of the collapse; however, eventually, there is a transition into positive values, and afterward, it is never violated. This transition is a signature of a standard collapsing scenario, where the masses accumulated around the central shell eventually negate an effective negative pressure component and compel all the matter to fall into a singularity.

\section{Formation of Null Surface}
We want to check if in general, a self-similar metric with a conformal structure in the PGL form can exhibit a minima on a space-like closed two-dimensional surface of minimum area and behave like a wormhole throat. To determine the throat conditions, we construct an embedding geometry for the general class of metric, having a conformal structure
\begin{equation}\label{metricgeodesic2}
ds^{2} = C^{2} \Bigg[A(r)^{2}dt^{2} - B(r)^{2}dr_{c}^{2} - r_{c}^{2}d\Omega^2 \Bigg].
\end{equation}

Since we are going to work on a spatial slice of constant $t$, the self-similar variable ($z = \frac{t}{r}$) should behave as a function of $r$ alone. On the constant time spatial slice, with $\theta=\pi/2$ the metric looks like
\begin{equation}\label{3_d}
dl^2 = B^{2}C^{2}dr_{c}^{2} + r_{c}^{2}C^{2}d\phi^{2}.
\end{equation}
$dl^2$ is the metric on a surface of revolution $\rho = \rho(z)$ embedded in a three-dimensional space with an Euclidean metric
\begin{equation}\label{R3d}
dl^2 = dZ^2 + d\rho^2 + \rho^2 d\phi^2,
\end{equation}
where $Z$, $\rho$ and $\phi$ are cylindrical coordinates. Comparing Eqs. (\ref{3_d}) and (\ref{R3d}) we get

\begin{eqnarray}\label{t1}
&&\rho^2 = r_{c}^{2}C^{2},\\&&\label{t2}
dZ^{2} + d\rho^{2} = B^{2}C^{2}dr_{c}^{2}.
\end{eqnarray}

For a constant $t$, $C(r_{c},t) = C_{0} f(r_{c})$ and we can write
\begin{equation}\label{t3}
d\rho = C dr_{c} + r_{c}dC = \left[ C_{0}f(r_{c}) + r_{c}C_{0}g(r_{c}) \right],
\end{equation}
where $g(r_{c}) = \frac{df}{dr_{c}}$. Using Eqs. (\ref{t1}), (\ref{t2}) and (\ref{t3}), we find

\begin{equation}
\frac{d\rho}{dZ} = \left[ \frac{(C_{0}f(r_{c}) + r_{c}C_{0}g(r_{c}))}{\left\lbrace C_{1}^{2}C_{0}^{2}f_{1}^{2}f^2 - (C_{0}f(r_{c}) + r_{c}C_{0}g(r_{c}))^2 \right\rbrace^{1/2}} \right].
\end{equation}

Note that we have used the metric condition $B = z^{n}C$. On a constant time slice this leads to
\begin{equation}
B = C_{1}f_{1}, ~ C_{1} = C_{0}t^{n}, ~ f_{1} = \frac{f(r_{c})}{r^n}.
\end{equation}

A throat (for a wormhole!) has the projected shape of a sphere, located at a certain value of the radial coordinates $r = r_{w}$. On an embedding diagram, this sphere of $r = r_{w}$ simply corresponds to a circle of radius $\rho$ on the surface of revolution. Naturally, at the throat, the radius of the circle $\rho(Z)$ should have a minimum. For this, the condition

\begin{equation}\label{ee4}
\frac{d\rho}{dZ}\Big|_{r_{w}} = 0,
\end{equation}
should be satisfied. This leads to a simple equation for the function $f(r_{c})$ leading to
\begin{equation}
f(r_{c}) = -r_{c}\frac{df}{dr_{c}}~ \Rightarrow f(r_{c}) = \frac{r_{0}}{r_{c}},
\end{equation}

at the throat, which is quite plausible. \\

The final fate of this collapse would likely be enveloped beyond an apparent horizon. The apparent horizon is a null surface defined by the condition,

\begin{equation}
g^{\mu\nu}Y_{,\mu}Y_{,\nu} = 0,
\end{equation}

where $Y(r,t)$ is the radius of the two-sphere. We investigate this condition of the formation of a null surface and find that for $C = C_{1} \frac{e^{\pm z^{-n}/n}}{z}$, it leads to
\begin{equation}
\left(1 \pm \frac{2}{z_{0}^n}\right)^{2} - \frac{4}{C_{1}^{2}} z_{0}^{(4-2n)} e^{\mp \frac{2}{n z_{0}^n}} \left(1 \pm \frac{1}{z_{0}^n}\right)^{2} = 0.
\end{equation}
We have assigned the value $z_{app}$ to $z$ when an apparent horizon might form. It is evident that at $z = z_{app}$, $t = t_{app}$ and $r = r_{app}$. We can calculate the time of formation of the apparent horizon for any particular shell labeled by $r$ from the above equation. Considering the fact that the collapse goes on as a continuous phase for a very long time, an apparent horizon is likely to form at a large value of time. We can make a reasonable approximation $\frac{r}{t} = \frac{1}{z} \Rightarrow 0$ and expand the terms in the above equation accordingly. The interesting outcome of the above equation is the fact that for all values of $n$, the collapse goes through the formation of multiple apparent horizons. For $n = 2$, the apparent horizon equation can be approximated into

\begin{equation}
1 \pm \frac{4}{z_{0}^2} - \frac{4}{C_{1}^2}\left(1 \mp \frac{1}{z_{0}^2}\right)\left(1 \pm \frac{2}{z_{0}^2}\right) = 0,
\end{equation}

which implies
\begin{equation}
z_{0}^2 = \pm \frac{4 (1 - C_{1}^{2})}{(C_{1}^{2} - 4)}.
\end{equation}

\section{Matching with an exterior geometry}

We assume the exterior region of the collapsing star to be a Schwarzschild metric
\begin{equation}\nonumber
ds^2 = (1-\zeta^2)dt_s^2 - \frac{dr^2}{1-\zeta^2} - r^2 d\Omega^2 ~,~ \zeta = \pm \sqrt{\frac{2m}{r}}.
\end{equation}
Taking $t_s = t + g(r)$ the metric can be transformed into
\begin{equation}
ds^2 = (1-\zeta^2)dt^2 \pm 2\zeta drdt - dr^2 - r^2 d\Omega^2, \label{PG}
\end{equation}
which is a PGL-compatible form provided
\begin{equation}
g^{\prime} = \pm \frac{\zeta}{1-\zeta^2} ~,~ g = \mp R \left( 2\sqrt{\frac{r}{R}} + \ln{\frac{\sqrt{r}-\sqrt{R}}{\sqrt{r}-\sqrt{R}}} \right).
\label{PGtSolution}
\end{equation}

In the present case, the interior solution and the Schwarzschild exterior do have a single PGL metric form as long as there is a continuity of
\begin{eqnarray}\label{GeneralFluidpsi}
\zeta = \left\{ \begin{array}{ll}
\mp \frac{\left(1 \mp (\frac{t}{r})^{n}\right)}{\frac{t}{r}}, \\
\mp \sqrt{\frac{R}{r}}.
\end{array} \right.
\end{eqnarray}
The continuity requirement leads to the equation
\begin{equation}
r^{\frac{3}{2}}\left\lbrace 1 \mp \left( \frac{t}{r}\right)^{n}\right\rbrace = \mp \sqrt{2m} t.
\end{equation}

Although this is a non-trivial continuity requirement, it is extremely suggestive of geodesic equations for an infalling particle. For instance, if $n < 0$, then at $t \Rightarrow \infty$ limit (where the singularity is supposed to develop), the above equation can be approximated into

\begin{equation}
r^{\frac{3}{2}} \pm \sqrt{2m} t = 0,
\end{equation}
which is exactly, the geodesic equation for a zero energy falling particle in the Schwarzschild exterior region. In other words, any freely falling particle towards the self-similar collapsing sphere hover at the infalling surface. During the collapse, if a stress develops on the boundary hyper-surface, due to the resulting surface tension the spherical shells can fall faster towards the singularity compared to a zero energy particle. Around $z \Rightarrow \infty$ ot $t \Rightarrow \infty$, the sphere approaches the singularity. The surface tension becomes negligible in this limit and the sphere behaves like a collapsing pressureless dust.

\section{Conclusion}
A considerable requirement remains for a \textit{complete} model of gravitational collapse within a proper geometric setup. The most intriguing point is that we do not know what will make a model complete in a physical as well as mathematical sense. At the outset, two requirements seem crucial : (i) the model must be able to describe the inner part of a black hole and free fall into it as part of the dynamic process, i.e., stellar collapse, and (ii) the model should not be limited to the usual diagonal metric form. This is where the Painlev\'e-Gullstrand-Lema\^itre form fits in nicely in this article. \\

We have described how a self-similar fluid distribution in GR can collapse due to it's own gravity. The self-similarity is preserved throughout the dynamic evolution. The exact solution we found has a non-diagonal form, written in a Painlev\'e-Gullstrand-Lema\^itre metric. The transition between a diagonal and a non-diagonal metric leads to a set of constraints on the metric tensor, ultimately leading to the exact solution. There is one free parameter in the exact solution, and depending on its parameter space, we discuss three possible outcomes. There is a possibility that the self-similar collapse will be followed by a bounce and dispersal of all the clustered matter distribution. The second possibility is that a rapid yet brief phase of collapse shall be followed by a bounce and another recollapse until a singularity is reached. The final possibility is that the sphere will experience a standard collapse and create a zero proper volume. The solution given in the manuscript unifies all these profiles in one exact form, given by $\frac{e^{-\frac{z^{-n}}{n}}}{z}$, where $n$ is the free parameter. We also present the profiles of Null energy conditions for all of the probable scenarios. It is found that a bounce is usually associated with a violation of the Null Energy Condition, while for a standard collapse, the energy conditions are never violated. The most curious property of the solution is that it can satisfy the condition of minima on a space-like closed two-dimensional surface, much like the null throat of a wormhole.

\section{Acknowledgments}
SC acknowledges Vellore Institute of Technology for the financial support through its Seed Grant (No. SG20230027), year 2023. CS thanks the Inter-University Centre for Astronomy and Astrophysics (IUCAA), Pune for financial support.

\end{document}